\documentstyle[12pt]{article}

\catcode`@=11

\def\beqra{\begin{eqnarray}} \def\eeqra{\end{eqnarray}}
\def\beqast{\begin{eqnarray*}} \def\eeqast{\end{eqnarray*}}
\def\beq{\begin{equation}}      \def\eeq{\end{equation}}
\def\be{\begin{enumerate}}   \def\ee{\end{enumerate}}

\def\fo{\hbox{{1}\kern-.25em\hbox{l}}}
\def\fnote#1#2{\begingroup\def\thefootnote{#1}\footnote{#2}\addtocounter
{footnote}{-1}\endgroup}

\def\sppt{Research supported in part by the
Robert A. Welch Foundation and NSF Grant PHY 9009850}

\def\ul{\underline}

\def\ch{\@startsection{section}{1}{\z@}{-3ex plus-1ex minus-.2ex}%
        {2ex plus.2ex}{\large\sc}}

\def\con{\ifmmode \hbox{\bf*} \else{\bf*}\fi}   % conjugation
\def\scon{\ifmmode \hbox{\footnotesize\rm\bf*} \else{\footnotesize\rm\bf*}\fi}

\def\0#1{\relax\ifmmode\mathaccent"7017{#1}%    % puts a little circle atop,
        \else\accent23#1\relax\fi}              % as a halo of a saint

\catcode`@=12

\def\eslash{\not{\hbox{\kern-2pt $E$}}}
\def\ans{{\it ansatz} }
\begin{document}

\hfill{UTTG-5-92}

\hfill{March 1992}

\vfill
\vspace{24pt}
\begin{center}

{\bf A New Ansatz for Quark and Lepton Mass Matrices\fnote{*}{\sppt}}

\vspace{24pt}

Gian F. Giudice

\vspace{8pt}

{\it Theory Group\\ Department of Physics\\
University of Texas\\ Austin, Texas 78712}

\vspace{36pt}

\ul{ABSTRACT}
\end{center}
\baselineskip=14pt

A new \ans for quark and lepton mass matrices is proposed in
the context of supersymmetric grand unified theories. The 13
parameters describing fermion masses and mixings are determined
in terms of only 6 free parameters, allowing 7 testable
predictions. The values of $V_{us}$, $V_{cb}$, $V_{ub}$, $m_u$,
$m_d$, $m_s$, and $m_b$ are then predicted as a function of the
3 charged lepton masses, $m_c$, $m_t$, and $\tan \beta$, the
ratio of Higgs vacuum expectation values. In particular the
Cabibbo angle and $m_s/m_d$ are determined in terms of only
lepton masses. All predictions are in very good agreement with
experiments.
\vfill
\pagebreak
\setcounter{page}{1}

One of the main open problems in particle physics is the understanding
of the quark and lepton mass matrices. In the Standard Model,
the fermion masses and mixings
are described by 13 free parameters. Attempts to compute these 13
phenomenological parameters within the framework of extensions of
the Standard Model so far have not proven very successful.
A less ambitious approach is to assume a particular form of the
fermion mass matrices, which allows a description of the physical
observables in terms of fewer parameters and which therefore makes
some testable predictions. It is hoped that this \ans hints at
the necessary symmetries of the interactions responsible for
fermion mass generation and then leads to the construction of a successful
theory beyond the Standard Model.

In this paper, I will present a new \ans for quark and lepton
mass matrices which involves only 6 free parameters, leading to
7 testable predictions. I begin by defining the
energy scale at which the \ans holds. If the \ans has anything
to do with the symmetries of an underlying theory of fermion
mass generation, then it should hold at the energy scale where
the extended theory breaks down into the Standard Model. The
subsequent running of the mass matrices from this energy scale
to the weak scale, according to the renormalization group equations,
may spoil the simplicity of the \ans and hide its symmetry
relations, but, it is hoped, provides the correct values of the
fermion masses and mixings.

I will assume here that the underlying theory is some kind of
Grand Unified Theory (GUT). Given the accurate measurements of
$\alpha_s$ and $\sin^2\theta_W$ at LEP, it is now clear that the
simplest way to achieve grand unification is to consider
a supersymmetric particle spectrum \cite{gut}. I will therefore
assume the minimal supersymmetric model below the GUT scale
$M_X$, and an unspecified GUT theory above $M_X$. Here I closely
follow the strategy proposed in ref.\cite{dhr}, where a different
\ans , that of
Georgi-Jarlskog \cite{geo}, is investigated.

Let us write the fermion mass terms, after spontaneous breaking
of the electroweak symmetry, as:
\beq
{\cal L}_{mass}={\bar q}^i_LU_{ij}u^j_R{v \over \sqrt{2}} \sin \beta
+{\bar q}^i_LD_{ij}d^j_R{v \over \sqrt{2}} \cos \beta
+{\bar l}^i_LE_{ij}e^j_R{v \over \sqrt{2}} \cos \beta + {\rm h.c.},
\eeq
where $i,j=1,2,3$ are generation indices, $\tan \beta$ is the ratio
of Higgs vacuum expectation values, and $v=246$ GeV. The \ans
proposed here for the Yukawa couplings is that, at the scale
$M_X$, $U$, $D$, and $E$ are Hermitian matrices of the form:
\beq
U=\pmatrix{0 & 0 & b \cr 0 & b & 0 \cr b & 0 & a \cr}, ~~~
D=\pmatrix{0 & f & 0 \cr f^* & d & d' \cr 0 & d'^* & c \cr}, ~~~
E=\pmatrix{0 & f & 0 \cr f^* & -3d & d' \cr 0 & d'^* & c \cr}.
\eeq

The relation between $D$ and $E$ is natural in GUT models where the
down quarks and the charged leptons lie in the same multiplet. For
instance, the coupling of a Higgs boson in the
{\bf 5} of SU$_5$ (or {\bf 10} of
SO$_{10}$) gives certain entries in the Yukawa coupling matrices of
the form $D_{ij}=E_{ij}$, while a Higgs boson in the
{\bf 45} of SU$_5$ (or {\bf 126} of
SO$_{10}$) gives $-3D_{ij}=E_{ij}$ \cite{geo}.
I will also assume $d'=2d$. This is
taken here as a purely phenomenological assumption, but
it is hoped that it
will find a group theoretical explanation,
analogous to the idea proposed in ref.\cite{geo},
in a complete GUT involving some generation symmetry.
I want to stress once again that my goal here
is to find simple relations that give successful phenomenological
predictions, which can be used as a guide for constructing
realistic models, rather than to find an explicit realization
of the \ans .
Using the freedom to redefine the fermion phases,
the \ans of eq.(2) now becomes:
\beq
U=\pmatrix{0 & 0 & b \cr 0 & b & 0 \cr b & 0 & a \cr}, ~~~
D=\pmatrix{0 & fe^{i\phi} & 0 \cr fe^{-i\phi} & d & 2d \cr
0 & 2d & c \cr}, ~~~
E=\pmatrix{0 & f & 0 \cr f & -3d & 2d \cr 0 & 2d & c \cr}.
\eeq
In eq.(3), $a,b,c,d,f,\phi$ are 6 real parameters which will
be used to compute the fermion masses and mixings, thus providing us
with 7 predictions.
Notice that without the requirement that $U$, $D$, and $E$ are
Hermitian, one would find two independent phases in eq.(3). As an
alternative to the hermiticity condition, I could have imposed
different relations among the phases in order to get rid of the
extra parameter.

The \ans in eq.(3) has to be run down to the weak scale.
The Yukawa coupling constants satisfy the one-loop renormalization group
equations:
\begin{eqnarray}
16 \pi^2 {d \over dt}U&=&(3UU^\dagger +DD^\dagger +3TrUU^\dagger -G_U)U,
\\
16 \pi^2 {d \over dt}D&=&(3DD^\dagger +UU^\dagger +3TrDD^\dagger
+TrEE^\dagger -G_D)D,
\\
16 \pi^2 {d \over dt}E&=&(3EE^\dagger +3TrDD^\dagger
+TrEE^\dagger -G_E)E,
\end{eqnarray}
where $t=\log \mu$, and $\mu$ is the renormalization scale, and
\beq
G_a=\sum_{i=1}^3c_a^ig_i^2,~~~~~~~~~~~~~~~~a=U,D,E.
\eeq
Finally, the $g_i$ are the gauge coupling constants which satisfy the
one-loop renormalization group equations:
\beq
16\pi^2{d\over dt}g_i=b_ig_i^3,
\eeq
and the coefficients $c_a^i$ and $b_i$ for the minimal supersymmetric
model are given in table 1.

In order to solve eqs.(4-6), I first redefine the quark fields so
that the Yukawa matrices transform as:
\beq
U\to K^\dagger UK\equiv U',~~~~D\to K^\dagger DK\equiv D',
\eeq
with $K$ chosen such that $U'$ is diagonal.
Neglecting in eqs.(4-6) all non-leading terms in Yukawa couplings
different from that of the top quark,
I find:
\begin{eqnarray}
16\pi^2{d\over dt} U'&=&\left[ 3{U'}_{33}^2 \pmatrix{0&&\cr&0&\cr&&1\cr}
+3{U'}_{33}^2-G_U\right] U',
\\
16\pi^2{d\over dt} D'&=&\left[ {U'}_{33}^2 \pmatrix{0&&\cr&0&\cr&&1\cr}
-G_D\right] D',
\\
16\pi^2{d\over dt} E&=&-G_EE.
\end{eqnarray}
The solutions of eqs.(10-12) are:
\begin{eqnarray}
U'(t)&=&\gamma_U \pmatrix{\xi^3&&\cr&\xi^3&\cr&&\xi^6\cr}U'(t_0),
\\
D'(t)&=&\gamma_D \pmatrix{1&&\cr&1&\cr&&\xi\cr}D'(t_0),
\\
E(t)&=&\gamma_E E(t_0),
\end{eqnarray}
where
\beq
\gamma_a=\exp \left( -{1\over 16\pi^2}\int_{t_0}^t dt G_a \right),~~~
\xi =\exp \left( {1\over 16\pi^2}\int_{t_0}^t dt {U'}^2_{33} \right),
\eeq
and the initial condition is $t_0=\log M_X$.
These integrals can be performed with the help of eq.(8) to obtain:
\beq
\gamma_a=\prod_{i=1}^{3}\left[ {g_i(t_0) \over g_i(t)}\right]^{c_a^i/b_i},
\eeq
\beq
\xi=\left[ 1+{3\over 4\pi^2}I{U'}^2_{33}(t_0)\right]^{-1/12}
=\left[ 1-{3\over 4\pi^2}{I\over \gamma_U^2}{U'}^2_{33}(t)\right]^{1/12},
\eeq
where $I=-\int_{t_0}^t dt \gamma_U^2$.

Thus the Yukawa coupling matrices renormalized at the weak scale $\mu$ are
given by:
\begin{eqnarray}
U_R&=&\gamma_U \pmatrix{\xi^3&&\cr&\xi^3&\cr&&\xi^6\cr}K^\dagger UK,
\\
D_R&=&\gamma_D \pmatrix{1&&\cr&1&\cr&&\xi\cr}K^\dagger DK,
\\
E_R&=&\gamma_E E,
\end{eqnarray}
where $U,D,E$ are the matrices at $M_X$ given by the \ans in eq.(3).
The matrix $K$, defined in eq.(9), is given by:
\beq
K=\pmatrix{\cos \theta & 0 & \sin \theta \cr 0 & 1 & 0 \cr -\sin \theta
& 0 & \cos \theta \cr}~~~~~~~~~\tan 2 \theta = {2b\over a}.
\eeq
Since $U_R$ is already diagonal, the Cabibbo-Kobayashi-Maskawa matrix is
simply given by the matrix $V$,
such that $V^\dagger D_R D_R^\dagger V$ is diagonal.

In order to diagonalize eqs.(20-21),
I use the fact that, because of the observed hierarchy of fermion
masses, the parameters in eq.(3) must satisfy $a>>b$, $c>>d>>f$.
Then I express $a,b,c,d,f$ in terms of $m_e$, $m_\mu $, $m_\tau $,
$m_c$, $ m_t$. After diagonalization, I find that the
Cabibbo-Kobayashi-Maskawa matrix becomes
\beq
V=\pmatrix{
c_1c_3e^{i\phi}-s_1s_2s_3 & s_1c_3e^{i\phi}+c_1s_2s_3 & -c_2s_3 \cr
-s_1c_2 & c_1c_2 & s_2 \cr
c_1s_3e^{i\phi}+s_1s_2c_3 & s_1s_3e^{i\phi}-c_1s_2c_3 & c_2c_3 \cr},
\eeq
where $c_1\equiv \cos \theta_1,s_1\equiv \sin \theta_1$,
etc. The 7 predictions
of the \ans in eq.(3) are:
\beq
V_{us}=|s_1|=3\sqrt{m_e \over m_\mu}\left( 1-{25\over 2}{m_e \over m_\mu}
+{16 \over 9}{m_\mu \over m_\tau}\right)
\eeq
\beq
V_{cb}=|s_2|={2\over 3}\xi^{-1}{m_\mu \over m_\tau}\left(
1-{m_e \over m_\mu}-{1 \over 9}{m_\mu \over m_\tau}\right)
\eeq
\beq
V_{ub}=|s_3|={\xi^2 \over \eta_c}{m_c\over m_t}
\eeq
\beq
m_u=\xi^3{\eta_u \over \eta_c^2}{m_c^2 \over m_t}
\eeq
\beq
m_d={\gamma_D \over \gamma_E} \eta_d 3 m_e \left(
1-8{m_e \over m_\mu}+{16 \over 9}{m_\mu \over m_\tau}\right)
\eeq
\beq
m_s={\gamma_D \over \gamma_E} \eta_s  {m_\mu \over 3} \left(
1+8{m_e \over m_\mu}-{16 \over 9}{m_\mu \over m_\tau}\right)
\eeq
\beq
m_b={\gamma_D \over \gamma_E} \xi \eta_b m_\tau
\eeq
where $\xi=[1-(3m_t^2I)/(2\pi^2 \gamma_U^2v^2\sin^2 \beta )]^{1/12}$.
Eqs.(24-30) have been obtained by diagonalizing eqs.(20-21), keeping
the first order corrections in $m_e/m_\mu$ and
$m_\mu/m_\tau$\footnote{I have checked that higher order terms
give corrections in eqs.(24-30) always smaller than $10^{-2}$.
Notice that terms of order
$m_\mu^2/m_\tau^2$ always contribute much less than terms of order
$m_e/m_\mu$, since the expansion is
really in $m_e/m_\mu$ and $m_\mu /3m_\tau$.}.
The effects of QCD renormalization of the quark mass $m_q$ from
the energy scale $\mu$ to the energy scale $\Lambda_q$, at which
the input quark mass is given, is contained in
$\eta_q \equiv m_q(\Lambda_q)/m_q(\mu )$. This means that each
quark mass appearing in eqs.(24-30) is defined at the energy scale
$\Lambda_q$; I choose
$\Lambda_q =1$ GeV for the three light quarks, and
$\Lambda_q=m_q$ for the three heavy quarks.
The renormalization
scale $\mu$ is taken to be equal to $m_t$; also the supersymmetry breaking
scale is taken to be equal to $\mu$ (=$m_t$), and threshold effects
due to the supersymmetric particle spectrum have been neglected.
The numerical predictions following from eqs.(24-30),
as a function of $m_t$ and $\xi$ (or, equivalently, $m_t$
and $\tan \beta$) are contained
in table 2, with the details of the input values illustrated
in the table caption.

A correct prediction of the bottom quark mass
constrains\footnote{Here and in table 2 I have neglected
the uncertainties in the determination of the quark mass due to the
$\mu$ (=$m_t$) dependence of the $\eta$'s and
$\gamma$'s. Such uncertainties, which amount to a few percent,
can be eliminated by computing the explicit
dependence on $m_t$, but nevertheless they cancel in the predictions for
the ratios of quark masses.}
$\xi =0.81 \pm 0.02$,
and therefore the ratio of Higgs vacuum
expectation values must satisfy:
\beq
\sin \beta \simeq {m_t\over 180~{\rm GeV}}.
\eeq
The condition $\tan \beta >1$, usually required by the electroweak
breaking mechanism in supersymmetric models, provides a lower bound
for the top quark mass of about 125 GeV. An upper bound of about 170 GeV
is obtained by requiring that the prediction for $m_u$ is consistent
with the result from chiral perturbation theory and
QCD sum rules ($m_u=5.1 \pm 1.5$ MeV \cite{Leut}), after errors on $m_c$
and $\xi$ have been taken into account. Note that because of the upper bound
on $m_t$, eq.(31) gives $\tan \beta < 3$ and therefore the approximation
of neglecting the Yukawa coupling for the bottom quark in the solution of
the renormalization group equations is justified.

As it is apparent from table 2, all the predictions are in good
agreement with the experimental results, for $\xi =0.81$ and
$m_t$ in the range 125--170 GeV. In particular, the Cabibbo
angle, which has been measured to one part in a hundred,
and $m_s/m_d$, which is precisely determined
from second order chiral perturbation theory \cite{Leutnucl},
are successfully predicted in terms of only lepton masses.

Notice that the phase $\phi$ does not appear in eqs.(24-30). The
CP violation effects of the Cabibbo-Kobayashi-Maskawa matrix are
then determined in terms of a new free parameter, $\phi$. The
parametrization-invariant CP violating quantity
\beq
|J|=|{\rm Im}(V_{ij}V_{lk}V^*_{ik}V^*_{lj})| ~~~~~~~{\rm for~~ any}~~
i\ne l,~~ j \ne k
\eeq
is given by
\beq
|J|=c_1c_2^2c_3s_1s_2s_3|\sin \phi |={\bar \xi}~ {130~{\rm GeV}\over
m_t} ~3.7 \cdot 10^{-5} |\sin \phi |,
\eeq
where the numerical values from table 2 have been used and
${\bar \xi} \equiv \xi/ 0.81$.

In conclusion, I have proposed a new \ans for quark and lepton
mass matrices, remnant of some unspecified supersymmetric GUT,
much in the same spirit of ref.\cite{dhr}. This \ans , eq.(3),
involves 6 free parameters and therefore leads to 7 predictions.
These are listed in eqs.(24-30) and compared with experimental
results in table 2. The striking agreement between predictions
and experiment suggests that eq.(3) may have something to do with
the theory of fermion mass generation.

\bigskip
I wish to thank Stuart Raby and Lawrence Hall for useful discussions.

\newpage
\begin{table}[t]
\begin{center}
\caption{The coefficient $c_a^i$ and $b_i$ for the minimal supersymmetric
model.}
\vspace{2em}
\begin{tabular}{c|ccc}
$i=$ & 1 & 2 & 3 \\
\hline
&&&\\
$c_U^i$ & $13 /15$ & 3 & $16 /3$ \\
$c_D^i$ & $7 /15$ & 3 & $16 /3$ \\
$c_E^i$ & $9 /5$ & 3 & 0 \\
&&&\\
\hline
&&&\\
$b_i$  & $33 /5$ & 1 & $-3$ \\
\end{tabular}
\end{center}
\end{table}

\begin{table}[t]
\begin{center}
\caption{The predictions of the \ans of eq.(3) compared with
experimental results. I have taken $M_X=1\cdot 10^{16}$ GeV and
$\alpha_X^{-1}= 25.1$, as predicted by supersymmetric grand
unification. For a range of $\mu$ between 125 and 170
GeV, I find $\gamma_U=3.2$,
$\gamma_D/\gamma_E=2.1$, $I=113$, $\eta_b=1.4$,
$\eta_c=1.8$, $\eta_{s,d,u}=2.0$. The masses of the 3 light quarks
are defined at 1 GeV, and the masses of the 3 heavy quarks are defined
at an energy scale equal to their masses. As input, I have taken $m_c=1.27
\pm 0.05$ and the charged lepton masses given in ref.[6].
The errors shown for the predictions take into account only the
uncertainty on $m_c$.
Finally ${\bar \xi}\equiv \xi / 0.81$.}
\vspace{2em}
\begin{tabular}{|c|c|c|c|}
  & prediction & experiment & reference \\*[1.mm]
\hline
\hline
&&& \\*[-3.5mm]
$V_{us}$ & 0.218 & $0.221 \pm 0.003$ & @ 90\% CL [6] \\*[1.mm]
\hline
&&& \\*[-3.5mm]
$V_{cb}$ & ${\bar \xi}^{-1}~~ 4.8\cdot 10^{-2}$ & $(4.4 \pm 0.9)$
$10^{-2}$ & [6] \\*[1.mm]
\hline
&&& \\*[-3.5mm]
$V_{ub}$ & ${\bar \xi}^2~~{130~ {\rm GeV}\over m_t}~~(3.6\pm 0.1$) $10^{-3}$
& $(4 \pm 3)$ $10^{-3}$ & @ 90\% CL [6] \\*[1.mm]
\hline
&&& \\*[-3.5mm]
$V_{ub}/V_{cb}$ & ${\bar \xi^3}~~{130~{\rm GeV}\over m_t}~~
(7.5\pm 0.3$) $10^{-2}$
& $(9 \pm 4)$ $10^{-2}$ & [6] \\*[1.mm]
\hline
&&& \\*[-3.5mm]
$m_u$ & $\bar{\xi}^3~{130~{\rm GeV}\over m_t}~ (4.1 \pm 0.3)$ MeV & $
5.1 \pm 1.5$ MeV & QCD sum rules [4] \\*[1.mm]
&&$4.5 \pm 1.4$ MeV & SU(4) mass relations [4] \\*[1.mm]
\hline
&&& \\*[-3.5mm]
$m_d$ & 6.9 MeV & $8.9 \pm 2.6$ MeV & QCD sum rules [4] \\*[1.mm]
&&$7.9 \pm 2.4$ MeV &  SU(4) mass relations [4] \\*[1.mm]
\hline
&&& \\*[-3.5mm]
$m_s$ & 138 MeV & $175 \pm 55$ MeV & QCD sum rules [4] \\*[1.mm]
&&$155 \pm 50$ MeV &  SU(4) mass relations [4] \\*[1.mm]
\hline
&&& \\*[-3.5mm]
$m_b$ & ${\bar \xi}~~ 4.25$ GeV & $4.25 \pm 0.10$ GeV & [4] \\*[1.mm]
\hline
&&& \\*[-3.5mm]
$m_u/m_d$ & $\bar{\xi}^3~~{130~ {\rm GeV}\over m_t}~~ (0.59 \pm 0.05)$ &
$ 0.56 \pm 0.08$ & chiral pert. theory [5] \\*[1.mm]
\hline
&&& \\*[-3.5mm]
$m_s/m_d$ & 20 & $20 \pm 2$ & chiral pert. theory [5] \\*[1.mm]
\hline
&&& \\*[-3.5mm]
$m_t$ & 125 -- 170 GeV & $140 \pm 35$ GeV   & fit of LEP data [7] \\*[1.mm]
&&$120^{~~+~27}_{~~-~28}$ GeV & fit of LEP data [8]  \\*[1.mm]
\hline
\end{tabular}
\end{center}
\end{table}

\end{document}